\documentclass[12pt]{article}
\usepackage{graphicx}

\newcommand{\const}{{\rm const}}
\newcommand{\var}{{\rm var}}
\newcommand{\cov}{{\rm cov}}
\newcommand{\diag}{{\rm diag}}
\newcommand{\rank}{{\rm rank}}
\newcommand{\sinc}{{\rm sinc}}

\begin{document}

\title{Statistical approach to linear inverse problems}

\author{V.~Yu.~Terebizh\thanks{\,valery@terebizh.ru}\\
\small{\textit{Crimean Astrophysical Observatory, 298409 Nauchny, Ukraine}}\\ 
\small{\textit{Institute of Astronomy RAN, Moscow 119017, Russian Federation}}\\  
 \small{\textit{ }}}

\date{\small{May 02, 2017}}

\maketitle

\begin{abstract}
The main features of the statistical approach to inverse problems are described 
on the example of a linear model with additive noise. The approach does not use 
any Bayesian hypothesis regarding an unknown object; instead, the standard 
statistical requirements \textit{for the procedure} for finding a desired 
object estimate are presented. In this way, it is possible to obtain stable and 
efficient inverse solutions in the framework of classical statistical theory. 
The exact representation is given for the \textit{feasible region} of inverse 
solutions, i.e., the set of inverse estimates that are in agreement, in the 
statistical sense, with the data and available \textit{a priory} information.
The typical feasible region has the form of an extremely elongated hole ellipsoid, 
the orientation and shape of which are determined by the Fisher information matrix. 
It is the spectrum of the Fisher matrix that provides an exhaustive description of 
the stability of the inverse problem under consideration. The method of 
constructing a nonlinear filter close to the optimal Kolmogorov--Wiener filter is 
presented. 
\end{abstract}


\section*{1.~Introduction}

In the usual sense, the \textit{inverse problem} is to find the \textit{object} 
$x_0$ from equation
$$
  y_0 = Hx_0,
  \eqno(1)
$$
where the \textit{image} $y_0$ describes the measured data, and $H$ is a 
known procedure (see, e.g., Tikhonov and Arsenin~[1977], Bertero~[1986]). 
We consider here the case when $H$ is a linear integral operator in the 
finite-dimensional space, so the unknown object $x_0$ and the observed image
$y_0$ can be treated as the ${n \times 1}$ and ${m \times 1}$ vectors,
respectively, while the \textit{point spread function} (PSF) $H$ is the ${m
\times n}$ matrix.

Inverse problems are especially characteristic for astronomy, which is still 
dealt predominantly with the interpretation of passive experiment (Feigelson 
and Babu~[2003, 2012]). In recent years, classical problems of this kind have 
been supplemented by tasks associated with the creation of an early Universe 
model based on microwave background measurements and study of distant galaxies.

In practice, even if the unique solution of the problem~(1) exists for any
$y_0$, i.e., the problem is \textit{well-posed}\footnote{For a linear problem, 
the continuity of the inverse mapping is a consequence of the stated
requirements.} in the sense of Hadamard~[1923], the solution can be strongly
\textit{unstable}. The latter term means that relative error propagation from
the image to the solution can be very large. Indeed, the data inevitably are
randomly noised, so, for a model with an additive noise, one should rewrite
equation (1) in the form 
$$
  y_0 = Hx_0+\xi,
  \eqno(2)
$$
 where $\xi$ is an {\em unknown} random noise pattern. Only the mean value of
noise, $a$, and its variance are known usually from the preliminary measurements.
Thus, if we shall try to minimize some kind of the \textit{misfit}, e.g. 
$||y_0-Hx-a||^2$, to find an appropriate inverse solution $x_*$, we obtain, as
a rule, the function with huge oscillations, because the \textit{least square}
(in general, the \textit{maximum likelihood}) solution $x_*$ is compelled to
`explain' sharp random noise fluctuations in the observed pattern $y_0$ purely 
by oscillations of the object's profile. In view of smoothing nature of the 
operator $H$, the amplitude of these oscillations should be large.

One can find in the literature descriptions a number of methods aimed to reach 
the stable inverse solutions (see, e.g., Press et.~al.~[1992], Jansson~[1997],
Evans and Stark~[2002]); the most widely applied now are the \textit{maximum
entropy} method (Janes~[1957a,b], Burg~[1967], Narayan and Nituananda~[1986])
and the \textit{regularization} method (Phillips~[1962], Tikhonov~[1963a,b],
Tikhonov and Arsenin~[1977]). Both approaches proceed from minimizing the sum
of two functionals:
$$
  x_\gamma = \arg \min_x \left[||y_0-Hx-a||^2 + \gamma \Phi(x)\right],
  \eqno(3)
$$
 where the first term, the misfit, measures the agreement of a trial
inverse solution $x$ to the model (2), while $\Phi(x)$, the \textit{stabilizing} 
or \textit{regularizing} functional, describes some kind of ``smoothness'' 
of the desired solution. The \textit{regularization parameter} $\gamma$ is 
introduced here to provide the trade-off between the accuracy and smoothness 
of the inverse solution.

It can be easily shown that the requirement (3) is equivalent to the Bayesian
way of estimation given \textit{a priory} information about the probabilistic 
ensemble of the allowable objects. The way is quite consistent and more efficient, 
comparing to the classic (``Simpsonian'', according to Eisenhart~[1964])
estimation, but only if we really have the needed \textit{a priory} information,
i.e., if $\Phi(x)$ is known. Since this happens comparatively rarely, some
intuitive forms of $\Phi(x)$ are usually applied, in particular, the quadratic 
norm
$$
  \Phi(x) = ||x||^2,
  \eqno(4)
$$
or more general norm in the Sobolev's space, or one of the (inequivalent to each 
other) `entropy' presentations. From the statistical point of view, this way 
corresponds to the introduction of the \textit{Bayes's hypothesis} that was many 
times criticized due to unavoidable subjectivity and inherent contradictions 
(see, e.g., Feller~[1957], Fisher~[1959], Rao~[1973], Cox and Hinkley~[1974], 
Szekely [1986]). The frequently stated dissatisfaction in relation to the 
Bayesian hypothesis prompted Press et.~al.~[1992], p.~808, rather figuratively 
express the moods: ``Courts have consistently held that academic license does 
not extend to shouting ``Bayesian''\, in a crowded hall.'' We note only that 
just the wide variety of forms the regularizing functional $\Phi(x)$ that were 
proposed for a same problem clearly shows the absence of a natural form of 
\textit{a priory} information about the searched object.

Meanwhile, it is possible to obtain the stable inverse solutions in a framework 
of the classic statistical theory. Some features of this way were described 
by Terebizh~[1995a,b, 2003, 2004]; in full extent the approach is presented in 
the book Terebizh~[2005], which also includes the statistical treatment of the 
other widely used inverse methods along with corresponding numerical algorithms. 
Both linear and non-linear models of data formation were considered; the first 
of them allows the quite general discussion, while the latter one is represented 
by the actual now \textit{phase problem} and the long-standing problem of the 
time series spectral estimation.

The main purpose of the below consideration is to give a brief outline of the
statistical approach within the linear data formation model with an additive 
noise. This model underlies more complicated cases and has very wide practical 
applications.

\section*{2.~Statistical formulation of the inverse problem}

Let us define the \textit{general linear model} by equations
$$
  \left\{   \begin{array}{ll}
  y_0 = Hx_0+\xi, & \\
  \langle \xi \rangle = a, \qquad
  \cov(\xi) \equiv \langle (\xi-a)(\xi-a)^T \rangle = C,
  \end{array}   \right.
  \eqno(5)
$$
where $n$-vector $x_0$ is an \textit{object}, the ${m \times n}$ PSF matrix $H$ 
is assumed to be known, as well as the observed \textit{image} $m$-vector $y_0$ 
(${m \ge n}$), and $\xi$ is the random \textit{noise} pattern with the known 
mean level $a$ and the positive definite covariance ${m \times m}$ matrix $C$. 
It is assumed that the original vectors have the form of columns; the angle 
brackets mean averaging on the probabilistic ensemble. Evidently, the density 
distribution $f(y|x_0)$ of the image is defined by the corresponding distribution 
of the noise. 

The sought inverse solution $\tilde x$ is considered as a statistical estimate
of the \textit{deterministic} object $x_0$ given its image, a PSF, properties
of the noise, and available \textit{a~priori} information about the object.
Being a function of the stochastic image $y_0$, the inverse solution $\tilde x$
is also a random vector, as a rule, with the mutually dependent components.
To define properly the notion of \textit{quality} of a trial estimate $x$, 
we should carry out two preliminary procedures. 

Firstly, correlations in the image~$y_0$ should be eliminated, because mutual 
correlations of the noise components $\{\xi_j\}_{j=1}^m$ does not allow applying 
the direct definition of the misfit in the form $||y_0-H\tilde x-a||^2$. 

Secondly, it is desirable to reduce the misfit vector dimension to the object's 
length~$n$. Indeed, it is possible to reach the quite good agreement of an image 
$y = Hx + \xi$ produced by some trial object $x$ with the observed image $y_0$ 
at the expense of good fit in the lengthy `wings' of the images, especially when 
${m \gg n}$. In fact, the wings represent mostly the noise patterns, while we 
have to fit primarily the part of the image that is caused by the smoothed object. 

The transition to the independent data set is based on the known linear 
transform
$$
  z_0 = C^{-1/2}(y_0-a), \qquad  \eta = C^{-1/2}(\xi-a),
  \qquad A = C^{-1/2}H,
  \eqno(6)
$$
 which converts the general model~(5) to the \textit{standard} model
$$
  \left\{   \begin{array}{ll}
  z_0 = Ax_0+\eta, & \\
  \langle \eta \rangle = 0, \qquad \cov(\eta) = E_m,
  \end{array}   \right.
  \eqno(7)
$$
 where $E_m$ is the unit ${m \times m}$ matrix. The matrix $C^{-1/2}$ in (6) 
is inverse to the square root of $C$; since the covariance matrix was assumed
positive definite, its spectrum is positive and the square root $C^{1/2}$
exists. Thus, we have now the relatively more simple linear model with an
additive white noise $\eta$ and the PSF matrix $A$.

The natural way to reach the second goal proceeds from the \textit{singular 
value decomposition} (SVD) of the matrix $A$ (see, e.g., Golub and Van 
Loan~[1989], Press et.~al.~[1992]). Assume that ${\rank(A) = n}$. Then
$$
  A = U \Delta V^T,
  \eqno(8)
$$
 where $U$ is an ${m \times n}$ column-orthogonal matrix, $\Delta$ is a 
diagonal ${n \times n}$ matrix with positive \textit{singular values} 
${\delta = [\delta_1,\ldots,\delta_n]^T}$ of $A$, placed in the order of 
their decrease, and an ${n \times n}$ matrix ${V = [v_1,\ldots,v_n]}$ is 
orthogonal:
$$
  U^TU = E_n, \qquad \Delta = \diag(\delta), \qquad V^{-1} = V^T.
  \eqno(9)
$$
 The corresponding decomposition of the object $x_0$ in the eigenvectors
system~$\{v_k\}$, namely
$$
  x_0 = V p_0 = \sum_{k=1}^n p_{0k}v_k, \qquad p_0 = V^T x_0,
  \eqno(10)
$$
 defines the vector $p_0$ of the \textit{object's principal components}. Like
the familiar Fourier coefficients, the principal components are often easier to
recover than the object itself. The multiplication of~(7) by $U^T$ is similar
to the application of the Fourier transform. Designating $n$-vectors 
$$
  \phi \equiv U^T z_0, \qquad \zeta \equiv U^T \eta,
  \eqno(11)
$$
 we obtain from (7) a final $n$-dimensional representation of the linear model:
$$
  \left\{   \begin{array}{ll}
  \phi = \Delta p_0+\zeta, & \\
  \langle \zeta \rangle = 0, \qquad \cov(\zeta) = E_n. &\\
  \end{array}   \right.
  \eqno(12)
$$
As was said above, the advantages of use of the `refined image' $\phi$ of
length~$n$ are especially appreciable when ${m \gg n}$.

\section*{3.~Feasible Region}

Assume, for simplicity, that the noise $\xi$ is a Gaussian deviate. Then
$\{\zeta_k\}$ in~(12) are independent Gaussian deviates with zero mean value
and unit variance, and the random variable $\|\phi-\Delta p_0\|^2 =
\sum_{k=1}^n \zeta_k^2$ has a $\chi^2$-distribution with $n$ degrees of freedom
(Cram\'er~[1946], Chapter~18). This result allows us to introduce a similar
random variable, namely the \textit{misfit}
$$
  \Theta(y_0|x) \equiv \|\phi-\Delta p\,\|^2\,,
  \eqno(13)
$$
 as a measure of the quality of a trial object's estimate ${x = Vp}$.

Let ${t_\gamma^{(n)} \ge 0}$ be a quantile of the $\chi_n^2$ distribution
$P_n(t)$, that is the root of equation ${P_n(t) = \gamma}$. Just as is usually
done in mathematical statistics (Cram\'er~[1946]), we shall choose the
appropriate boundary significance levels for an inverse solution $\alpha_1$ and
$\alpha_2$ ($0 \le \alpha_1 \le \alpha_2 \le 1$). By definition, a trial
object's estimate $x$ is called \textit{feasible}, if
$$
  t_{1-\alpha_2}^{(n)} \le \Theta(y_0|x) \le t_{1-\alpha_1}^{(n)}.
  \eqno(14)
$$
 We simply require of a feasible estimate $x$ that its image $y(x)$ should have
moderate deviation, in the statistical sense, from the observed image $y_0$.
Inequalities~(14) define the \textit{feasible region} (FR), consisting of all
the object's estimates $\{x\}$ that have feasible agreement with the data. It
is convenient to call $x$ \textit{the estimate of significance level}~$\alpha$,
if the misfit $\Theta(y_0|x) = t_{1-\alpha}^{(n)}$, that is
$$
 \|\phi-\Delta p\,\|^2 = t_{1-\alpha}^{(n)}.
 \eqno(15)
$$

According to the known Gauss--Markov theorem, the \textit{least squares
estimate} (LSE) 
$$
  x_* = (A^T A)^{-1}A^T z_0
  \eqno(16)
$$
 has the smallest variance of all the unbiased object's estimates (Lawson and
Hanson~[1974]). It follows from~(8) and~(16) that 
$$
  x_* = V p_* = \sum_{k=1}^n p_{*k}v_k, \qquad
  p_* = \Delta^{-1} \phi.
  \eqno(17)
$$
 Equations (17) define the \textit{principal components of LSE} $p_*$. Unlike
the object's principal components $\{p_{0k}\}$, the LSE components $\{p_{*k}\}$ 
are random variables. One can easily find the mean value and the covariance
matrix of the LSE: 
$$
  \langle p_* \rangle = p_0, \qquad \cov(p_*) = \Lambda^{-1},
  \eqno(18)
$$
 where the matrix 
$$
  \Lambda \equiv \Delta^2 = \diag(\lambda_1,\ldots,\lambda_n),
  \qquad \lambda_k = \delta_k^2.
  \eqno(19)
$$
 Thus, the LSE principal components $\{p_{*k}\}$ are the unbiased estimates of
$p_{0k}$, and ${\var(p_{*k}) = \lambda_k^{-1}}$. Usually, the `tail' of the
sequence $\{\lambda_k\}$ is very small, so the variance of corresponding
$\{p_{*k}\}$ and consequently the variance of LSE are huge.

Let us remind the geometrical interpretation of this phenomenon. With the help
of~(8) and~(17), it is easy to transform the definition~(15) into the form 
$$
  (x-x_*)^T I (x-x_*) = t_{1-\alpha}^{(n)}, \qquad
  I = A^T A = V \Lambda V^T.
  \eqno(20)
$$
 Therefore, the feasible region consists of hollow ellipsoids, centered at
the LSE, and the shape of ellipsoids is defined by the ${n \times n}$ matrix
$I$. The latter is a representation of the \textit{Fisher matrix} with the
components 
$$
  I_{ik}(x_0) \equiv \biggl\langle  \frac{\partial}{\partial
  x_{0i}}\, \ln f(y_0|x_0)\, \frac{\partial}{\partial
  x_{0k}}\, \ln f(y_0|x_0) \biggr\rangle,\, \quad
  i,k = 1,\ldots,n,
  \eqno(21)
$$
 for the particular inverse problem (5) under consideration
(Terebizh~[1995a,b]). The lengths of semi-axes of the FR ellipsoid are
determined by expressions $\ell_k = \sqrt{t_{1-\alpha}^{(n)}/\lambda_k}$. Small
values of the farthest eigenvalues $\{\lambda_k\}$ in the spectrum of
matrix~$I$ give rise to an extremely elongated shape for the~FR. Just that
phenomenon reveals itself in the well-known instability of inverse solutions.
Indeed, a trial object's estimate $x$ that situated very far from the true
object $x_0$ can produce the image $y$ that is in feasible agreement, in a
scale of natural noise fluctuations, with the really observed image $y_0$.

The feasible region usually does not include the LSE and the manifold in its
vicinity. Specifying the said in the Introduction, the reason is that the
object's estimates close to LSE try to `explain' all details of the observed
image, irrespective of their statistical significance. Since the model~(5)
supposes essential smoothing of the object, one should admit large erroneous
oscillations in the LSE in order to fit tiny random fluctuations in the image.
The formal base of the corresponding requirement is given by defining
\textit{two} significance levels $(\alpha_1,\alpha_2)$, as it is usually done
in mathematical statistics.

\section*{4.~Optimal linear filter}

It is possible to mitigate the harmful influence of the small eigenvalues of
the Fisher matrix by introducing into~(17) the appropriate set of weights ${w =
[w_1, \ldots, w_n]^T}$, so 
$$
  x_w \equiv \sum_{k=1}^n w_k p_{*k} v_k = VWp_*,
  \qquad W = \diag(w).
  \eqno(22)
$$
 A number of known inverse solutions, in particular, Kolmogorov~[1941] and
Wiener~[1942] \textit{optimal} estimate, the \textit{regularized} solution by
Phillips~[1962] and Tikhonov~[1963], and the \textit{truncated} estimate
(Varah~[1973], Hansen~[1987, 1993], Press et.~al.~[1992]), belong to the class
of linearly filtered estimates. It follows from~(10) and~(22) that the squared
error of the filtered estimate 
$$
  \varepsilon_w^2 \equiv  \langle \|x_w - x_0\|^2 \rangle =
  \sum_{k=1}^n \left[ w_k^2/\lambda_k +
  (1-w_k)^2 p_{0k}^2 \right].
  \eqno(23)
$$
 As one can see, the error is minimized by the set of weights 
$$
  \tilde w_k = \lambda_k p_{0k}^2/(1+\lambda_k p_{0k}^2)\,,
  \qquad k = 1,2,\ldots,n,
  \eqno(24)
$$
 which constitutes the \textit{optimal Wiener filter} $\tilde W(p_0) = \diag
\left[ \tilde w(p_0) \right]$. Consequently, the best of linearly filtered
estimates of the object is 
$$
  \tilde x_w = \sum_{k=1}^n \tilde w_k p_{*k} v_k =
  V \tilde p_w, \qquad \tilde p_w = \tilde W(p_0) p_*.
  \eqno(25)
$$

An important feature of the optimal filter is that the weights $\tilde w$
depend not only on the known properties of the PSF and the noise but also upon
the object itself. For that reason, the filter can be applied only in the
Bayesian approach to inverse problems. It is worth noting, in this connection,
that the investigations of Kolmogorov~[1941] and Wiener~[1942] focused on time
series analysis, where the Bayesian approach is well justified since the
Gaussian nature of ensembles is ensured by the central limit theorem. For most
other inverse problems, and in particular, image restoration, the availability
of both object ensembles and prior probability distributions on those ensembles
is unnatural.

We can simplify the general description of the FR for the linearly filtered
estimates by substituting ${p_w = Wp_*}$ into~(15) or (20). The result is: 
$$
  \|(W - E_n)\,\phi\,\|^2 = t_{1-\alpha}^{(n)}.
  \eqno(26)
$$
 This condition imposes restrictions on the system of weights $w$. Then~(22)
enables the filtered estimate to be found.

One can expect that the requirement~(26) is satisfied for the optimal filter
${W = \tilde W(p_0)}$ at moderate values of the significance level $\alpha$.
Indeed, extensive numerical simulations are in agreement with this assumption;
the corresponding significance level usually is more than $0.70$.

\section*{5.~Quasi-optimal filter}

If it were possible to find a good approximation of the object's principal
components $\{p_{0k}\}$ in~(24) with only the given and the observed
quantities, the corresponding filter would doubtless have a practical value,
but we have no \textit{a~priori} information for such immediate approximation.
At the same time, and that is the key point of the quasi-optimal filtering, we
have enough information about the \textit{structure} of the optimal estimate
$\tilde x_w$, in order to require similar properties for the estimate of the
object searched for.

By substituting $x_0$ from~(10) and $\tilde x_w$ from~(25) into~(23), and
noting that the orthogonal transform does not change the vector norm, we obtain 
$$
  \langle \|\tilde W(p_0)p_* - p_0\|^2 \rangle =
  \tilde \varepsilon_w^2(p_0).
  \eqno(27)
$$
 This equation simply gives another representation of the error of the optimal
filter, which, by definition, is smallest in the class of linear filters.

Let us now consider a trial estimate $p$ close to $p_0$ (Fig.~1). Taking into
account~(27), we shall require that the filter 
$$
  \tilde W(p) = \diag \left[ \tilde w(p) \right], \qquad
  \tilde w_k(p) = \lambda_k p_k^2/(1+\lambda_k p_k^2),
  \eqno(28)
$$
 which is based on such an estimate, had the minimal error: 
$$
  \langle \|\tilde W(p)p_* - p\,\|^2 \rangle = \min.
  \eqno(29)
$$
 Note that the \textit{quasi-optimal filter} (28) has the same structure as the
optimal Wiener filter~(24). {\em Thus, we search for the estimate that most
closely simulates behaviour of the best inverse solution}.

\begin{figure}[h]   
 \centerline{\includegraphics[width=17pc]{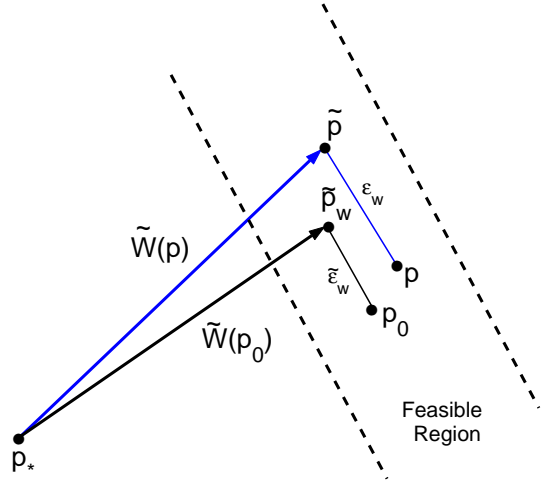}}
 \caption{{\footnotesize Schematic representation of the optimal and the
  quasi-optimal filtering in the space of principal components.
  $p_0$~-- object, $p_*$~-- Least Squares Estimate,
  $\tilde W(p_0)$~-- optimal filter, $\tilde p_w$~-- optimal
  estimate of the object, $p$~-- trial estimate, $\tilde W(p)$~--
  Wiener filter for the trial estimate, $\tilde p$~--
  quasi-optimal estimate of the object. The errors of the filters
  are shown by the segments $\tilde \varepsilon_w(p_0)$ and
  $\varepsilon_w(p)$.}}
\end{figure}

If we depart from the averaging procedure, which is executable only in theory,
and add the condition~(26), which requires that the trial object's estimate
belongs to the feasible region, we obtain the simultaneous conditions 
$$
  \left\{   \begin{array}{ll}
  \| [\tilde W(p)-E_n] \,\phi\,\|^2 = t_{1-\alpha}^{(n)}, & \\
  &\\
  \|\tilde W(p)p_* - p\,\|^2 = \min. &\\
  \end{array}   \right.
  \eqno(30)
$$
 The solution $p_{\min}$ of this system allows us to find the
\textit{quasi-optimal estimates} of the object and its principal components: 
$$
  \tilde p = \tilde W(p_{\min})\,p_*, \qquad
  \tilde x = V \tilde p.
  \eqno(31)
$$
 Indeed, we are ultimately interested not in the $p_{\min}$ that is intended to 
replace $p_0$ only in argument of the filter (see Fig.~1), but in the
\textit{filtered} estimate of the principal components~$\tilde p$, which is
analogous to the optimal Wiener estimate $\tilde p_w$ in~(25).

In the components of the corresponding vectors, equations (30) can be written
as 
$$
  \left\{   \begin{array}{ll}
  \sum_{k=1}^n \left[ \tilde w_k(p) -1 \right]^2 \phi_k^2 =
  t_{1-\alpha}^{(n)}, & \\
  & \\
  \sum_{k=1}^n \left[ \tilde w_k(p)\,p_{*k}-p_k \right]^2 =
  \min, &\\
  \end{array}   \right.
  \eqno(30')
$$
 where $\tilde w_k(p)$ are given by (28) and the vector $\phi=\{\phi_k\}$ was
defined by~(11).

Unlike the Wiener filter, the quasi-optimal filter is nonlinear with respect to
the LSE $p_*$, because a solution $p_{\min}$ of the system (30) is dependent
upon $p_*$, and then we should apply filtering according to~(31).

Since both functionals in equations~(30) are positive definite, and the second 
functional is non-degenerate, the solution of the constrained minimization
problem~(30) is unique (Press et.~al.~[1992], \S~18.4).

To understand the sense of the quasi-optimal filtering better, it is useful to 
bear in mind the following. The object and its least squares estimate were 
held fixed when searching for the optimal filter, whereas the filter structure 
has been optimized. On the contrary, equations~(28) and~(29) fix the previously
determined structure of the filter (and the LSE, of course), concentrating
attention on the search for an appropriate estimate of the object. Such 
an approach seems to be quite justified, because the simultaneous searches for 
both the best filter and the good inverse solution are possible only if complete
information about the object is available. The efficiency of the optimal
filtering should be high enough in the vicinity of the unknown object; so we 
do have reason to fix a form of the best filter for an estimate close to the
object.

\begin{figure}[h]   
 \centerline{\includegraphics[width=30pc]{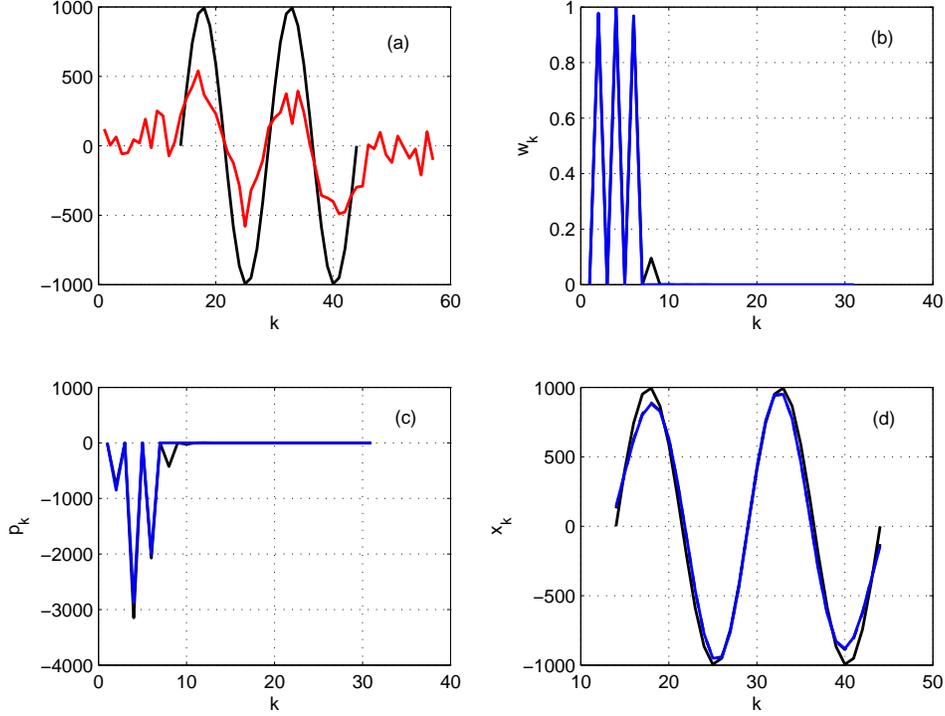}}
 \caption{{\footnotesize (a) Object (black), and its blurred image (red).
 (b) Weights of the optimal Wiener (black) and the
 quasi-optimal (blue) filters. (c) Principal components of the
 object (black), and the quasi-optimal estimate (blue). (d) The object
 (solid black), the optimal (dashed), and the quasi-optimal (blue)
 estimates. Blue color dominates when lines coincide.}}
\end{figure}

\subsection*{6. Model cases}

Equations (28), (30) and~(31) form the basis for an algorithm that can be
programmed with a high-level programming language. To show the distinction 
between two filters under discussion more clearly, we deliberately consider
here simple examples. 

\begin{figure}[h]   
 \centerline{\includegraphics[width=30pc]{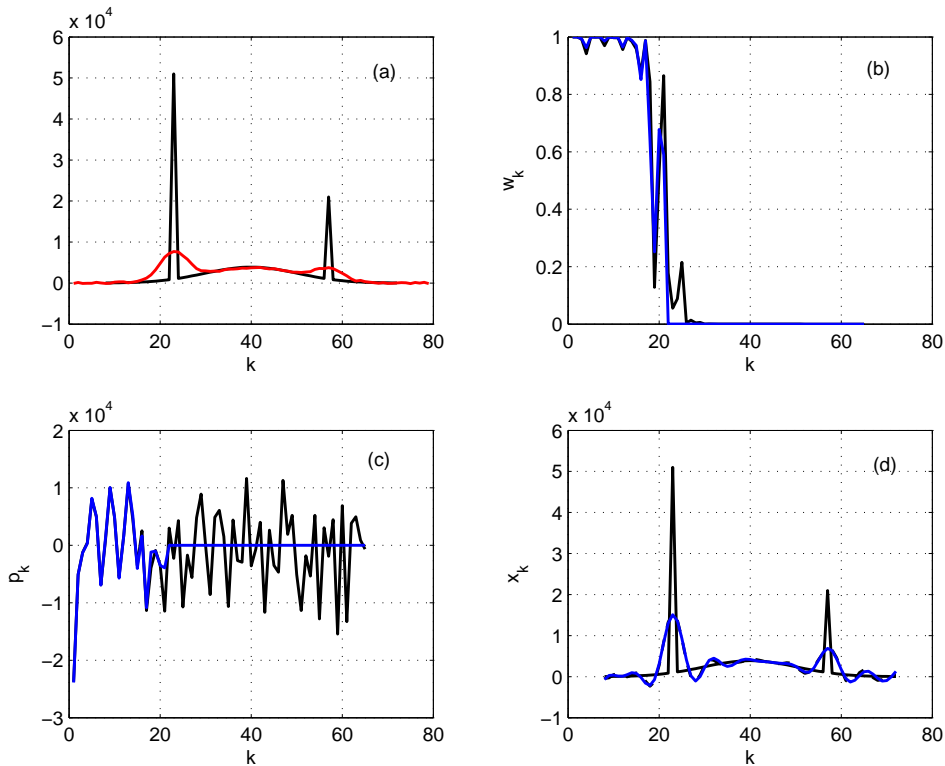}}
 \caption{{\footnotesize (a) Object (black), and its blurred image (red).
 (b) Weights of the optimal Wiener (black) and the
 quasi-optimal (blue) filters. (c) Principal components of the
 object (black), and the quasi-optimal estimate (blue). (d) The object
 (solid black), the optimal (dashed), and the quasi-optimal (blue)
 estimates. Blue color dominates when lines coincide.}}
\end{figure}

Figure~2 describes restoration of a low-frequency object that we have assumed
to be the portion of a sinusoid having an amplitude $1000$. A space-invariant
PSF 
$$
  h(t-t') = R^{-1}\,\sinc^2\left[(t-t')/R \right]
  \eqno(32)
$$
 was adopted, where ${\sinc(t) \equiv \sin(\pi t)/(\pi t)}$, and the
characteristic radius $R$ was taken as $9$ pixels. Function (32) can be
considered as the one-dimensional analogue of the Airy diffraction pattern. 
The mean level of the Gaussian white noise $a$ was taken as zero, its standard
deviation $\sigma_{\xi}$ as $100$. The significance levels of the filters were
equal each other.

As one can see from Figs.~2b and 2c, the quasi-optimal weight function and
the principal components are practically coincide with the corresponding
optimal values at low spatial frequencies. The same is true for the restored
objects; both estimates are indistinguishable in the scale of Fig.~2d. Note
the removal of the erroneous high-frequency oscillations in the object
estimates, and the non-monotonic behaviour of both the optimal and the
quasi-optimal weights, which is distinct from those for a \textit{truncated}
estimate. The latter leaves in the object's estimate simply a few of the first
principal components; the quasi-optimal filter leaves only those principal
components that have the highest accuracy of restoration. The errors of the
Wiener and the quasi-optimal filters were nearly the same for the considered
example.

Figure~3 depicts a traditionally difficult model case that incorporates
superposition of the sharp and smooth details. The Gaussian PSF has been
applied this time with the standard deviation ${\sigma_{PSF} = 3}$ pixels; the
noise standard deviation has remained as above. As one can see from Figure~3,
both the optimal and the quasi-optimal estimates have similar qualities.

We should not assert the claims to quality of restoration in the last case, 
because even the theoretically best filter, the Wiener's one, shows rather
inexpressive results in that case. Is is more important that the numerous 
model cases testify that the Wiener's and the quasi-optimal filters provide 
very close results. So to speak, a rogue is able to cure any disease, whereas 
a true physician is only able to do what is possible under the circumstances.

The discussion of the non-negativity condition and the Poisson model is given
elsewhere (Terebizh~[2003, 2005]).

\section*{7.~Concluding remarks}

It is appropriate to emphasize the importance of the Fisher matrix (21) that 
plays a fundamental role not only in the linear model but also in the general
inverse problem (Terebizh~[1995a,b]). To simplify the discussion, we assumed 
above that the spectrum of the matrix $I$ can include the arbitrarily small,
but strictly non-zero eigenvalues $\{\lambda_k\}$. This restriction is not
essential for the final results; the case of some zero eigenvalues can be
treated with the aid of the known additional procedure with the LSE (Press
et.~al.~[1992]).

To avoid misunderstanding, let us repeat once more that the Bayesian way of
estimation, in itself, is irreproachable. Moreover, since it incorporates an
additional \textit{a priory} information about the object, the quality of the
Bayesian estimation is higher of those in the classical approach. Insurmountable 
contradictions arise only at the reference to the so called \textit{Bayes's
hypothesis}, i.e., at substitution really available \textit{a priory}
information concerning the object by some speculative general principles.

Perhaps, the reasons are better visible at the analysis of the concrete
procedures. Examples concerning the \textit{maximum entropy} method were
considered by Terebizh~[2005], \S3.3; we shall touch here the regularization
procedure that is described more elaborately in Appendix~1.

In the Phillips--Tikhonov's approach, the most widely used stabilizing
requirement consists in condition of minimal `power' of the inverse solution,
namely 
$$
  \sum_{k=1}^n x_k^2 = \min,
  \eqno(33)
$$
 given of course the acceptable misfit of the observed and the trial images. We
have found out above an important role of the principal components ${p = \{p_k\}}$ 
that are associated with the estimate $x$ by the orthogonal transform 
${p = V^Tx}$. This role is based mainly on the mutual statistical independence 
of principal components, which allows to apply the theoretically most effective 
way of extracting information from an unstable least squares estimate: it is 
necessary to take principal components one after another according to descending 
order of the eigenvalues $\{\lambda_k\}$ of the Fisher matrix~$I$. As the $I$ 
spectrum covers an extremely wide range, the estimates of the principal components 
have the \textit{essentially different accuracy}. According to equations (18) and
(19), ${\var(p_{*k}) = \lambda_k^{-1}}$; for common in practice situations that
means tens of orders in value. Meanwhile, the invariance of a vector norm at the 
orthogonal transformation entails the equivalence of the condition (33) and the
requirement 
$$
  \sum_{k=1}^n p_k^2 = \min.
  \eqno(34)
$$
 Evidently, the direct summation the variables of essentially different
accuracy is not an optimal procedure. For example, it seems better to take
$p_k^2$ with some weights $g(\lambda_k)$ that are dependent on the corresponding
variance of the least squares estimate. This way leads to the condition 
$$
  \sum_{k=1}^n g(\lambda_k)\,p_k^2 = \min,
  \eqno(35)
$$
 which, however, assumes some subjective choice of the weight function.

The objectively justified requirement is given in (30) by the close on sense,
but more refined condition ${F(p)= \min}$, where 
$$
  F(p) = \|\tilde W(p)p_* - p\,\|^2.
  \eqno(36)
$$
 Taking into account (28), we can rewrite the above expression as 
$$
  F(p) = \sum_{k=1}^n \left( \frac{\lambda_k p_k^2}
  {1+\lambda_k p_k^2}\,p_{*k} - p_k \right)^2.
  \eqno(37)
$$
Usually, the production $\lambda_k p_k^2 \gg 1$ for the small $k$ and then
quickly decreases with the growth of $k$ to the values much less than 1. 
Therefore, the summation range in (37) can be approximately divided into 
two regions with the boundary value of $K$, such that 
${\lambda_K p_K^2 \simeq 1}$. Then 
$$
  F(p) \simeq \sum_{k=1}^K (p_k-p_{*k})^2 + \sum_{k=K+1}^n p_k^2.
  \eqno(38)
$$
 We see that the sought estimates $p_k$ and the LSE principal components
$p_{*k}$ should be close only for the large first eigenvalues $\lambda_k$ of
the Fisher matrix. Such a requirement is quite reasonable in view of high
accuracy of the first principal components of LSE. At the same time, just the
summary `tail' of $\{p_k^2\}$ of low accuracy is minimized when $\lambda_k
p_k^2 \ll 1$.

On the contrary, the condition (34) equally minimizes all principal
components irrelatively of their accuracy. As the numerous model cases show,
that entails too large systematic shift, the \textit{bias} of the regularized
inverse solution. In geometrical language it means that the point of a contact
the ellipsoidal feasible region and the `stabilizing' region noticeably depends
on the shape of the latter region.

From the viewpoint of the regularization theory, it might seem that the
functional (36) is a smoothing functional similar to $\|x\|^2$ or to one of the
several forms of the `entropy' ${\cal E}(x)$. Indeed, the condition ${F(p)=
\min}$ promotes stabilization of the inverse solution, but \textit{the origin}
of this functional is of vital importance. The Bayesian hypothesis proposes to
compensate the lack of \textit{a~priori} information by some general principle 
that directly concerns the properties of the sought object $x$ itself. Obviously, 
it is possible to offer an unlimited number of such principles. On the contrary, 
we rely on the intrinsic reserves of the inverse theory. The relatively much 
weaker assumption has been applied in the proposed above way, namely, that the 
optimal Wiener filter retains a high efficiency in the local vicinity of the 
unknown object. It appears that, instead of the prior information about 
\textit{the object}, it is enough to lay down some reasonable statistical 
requirements only \textit{to the restoration procedure}.

\section*{Appendix~1. Statistical treatment of the Phillips--Tikhonov
 regularization}

The statistical point of view on the Phillips--Tikhonov procedure is 
illustrated below by the example when regularizing functional $\Phi(x)$ 
is taken in the form~(4). As it was stated in the main text, the corresponding 
feasible region (FR) is defined by equation (15). Combining this condition with 
(4) and taking into account that the norm of the vector ${x = Vp}$\, remains  
the same under orthogonal transformation, we come to the simultaneous system 
$$
  \left\{   \begin{array}{ll}
  \|\phi - \Delta p\,\|^2 = t_{1-\alpha}^{(n)}, &
  \qquad \alpha_1 \le \alpha \le \alpha_2,\\
  \|p\,\|^2 = \min, &
  \end{array}   \right.
  \eqno(A1.1)
$$
 where vector $\phi$ and the diagonal matrix $\Delta$ are defined by equations 
(11) and (9), respectively. The first of the above conditions describes an 
ellipsoidal FR which elements, by definition, provide the feasible misfit between 
the observed and trial images. Let us consider elements situated on the sphere 
${\|p\,\|^2 = \const}$ of small radius centered at the origin. Gradually
enlarging the radius of the sphere, we take into consideration elements of the 
larger power, so the point of contact of the spheres family and the fixed FR
ellipsoid gives the Phillips--Tikhonov's inverse estimate.

According to the method of Lagrange multipliers, the necessary extremum
conditions follow from minimization of the auxiliary functional 
$$
  {\cal L}_{\gamma}(p) = \|\phi -\Delta p\,\|^2 + \gamma\,\|p\,\|^2,
  \eqno(A1.2)
$$
 where the scalar $\gamma \ge 0$ as yet is free. If the vector $p_{\gamma}$ that 
provides minimum of the Lagrange function is found for any $\gamma$, then the
desired value of the regularization parameter $\gamma$ is defined by
substitution $p_{\gamma}$ into the first of conditions (A1.1).

In order to find an explicit form of $p_{\gamma}$, let us rewrite (A1.2) as 
$$
  {\cal L}_{\gamma}(p) = \|\bar \phi - \bar \Delta p\,\|^2 + \const,
  \eqno(A1.3)
$$
 where 
$$
  \bar \Delta = (\Delta^2+\gamma E_n)^{1/2}, \qquad
  \bar \phi = \bar \Delta^{-1}\Delta \phi.
  \eqno(A1.4)
$$
 The diagonal matrix $\bar \Delta$ is of size ${n \times n}$, and $\bar \phi$ is
the $n$-vector. The minimum of the functional (A1.3) gives the element 
$$
  p_\gamma = \bar \Delta^{-1} \bar \phi =
  (\Lambda+\gamma\,E_n)^{-1} \Lambda p_*,
  \eqno(A1.5)
$$
 where $\Lambda = \Delta^2$, and $p_*=\Delta^{-1}\phi$ is the \textit{least square
estimate} (LSE) of principal components (see the first of equations (A1.1)). By
defining the diagonal ${n \times n}$ matrix 
$$
  W_{\gamma} \equiv (\Lambda+\gamma\,E_n)^{-1}\Lambda =
  \diag \left( \frac{\lambda_k}{\lambda_k+\gamma}
  \right)_{k=1}^n\,,
  \eqno(A1.6)
$$
 we can write down the regularized vector of the principal components and the
corresponding inverse solution as 
$$
  p_\gamma = W_\gamma p_*, \qquad X_\gamma = V p_\gamma.
  \eqno(A1.7)
$$
 Comparing (A1.6) and (A1.7) with the general definition (22) of a filtered
estimate, we come to conclusion that \textit{the regularized according to
Phillips and Tikhonov inverse solution belongs to the class of estimates that
were obtained by linear filtering of the least squares (maximum likelihood)
estimate}.

Remind that the definition of FR for a linear filtered estimate is given by
(26), or, in an unfolded form, 
$$
  \sum_{k=1}^n (1-w_k)^2 \phi_k^2 = t_{1-\alpha}^{(n)}.
  \eqno(A1.8)
$$
 Substituting here the weights according to (A1.6), we come to the following
equation for the inverse value of the regularization parameter ${\mu \equiv
1/\gamma}$: 
$$
  f(\mu) =  t_{1-\alpha}^{(n)},
  \eqno(A1.9)
$$
 where the function 
$$
  f(\mu) \equiv \sum_{k=1}^n \left( \frac{\phi_k}
  {1+ \mu \lambda_k} \right)^2, \qquad 0 \le \mu < \infty.
  \eqno(A1.10)
$$
According to (11) and (6), $\phi = U^Tz_0 = U^TC^{-1/2}(y_0-a)$, where $y_0$ is
the observed image.

As one can see, function $f(\mu)$ monotonously descends from ${f(0) =
\|\phi\|^2}$ down to zero when ${\mu \to \infty}$, so any of standard numerical
methods can be easily applied to find the unique root of equation (A1.9), i.e.,
the regularization parameter $\gamma$ (see, e.g., Booth~[1955], Press et
al.~[1992]). After that the formulae (A1.6) and (A1.7) allow us to find the
statistical estimate of the object of given significance level.

The above algorithm is intended for the restoration of any objects, both positive 
and with some negative components. If \textit{a priory} information about the
object $x_0$ assumes non-negativity of the all its components, we have, instead
of (A1.1), the following simultaneous system for estimating the vector ${x =
Vp}$: 
$$
  \left\{   \begin{array}{ll}
  \|\phi - \Delta p\,\|^2 = t_{1-\alpha}^{(n)}, &
  \qquad \alpha_1 \le \alpha \le \alpha_2,\\ 
  \|p\,\|^2 = \min,\\ 
  Vp \ge 0.
  \end{array}   \right.
  \eqno(A1.11)
$$
 It was shown by Terebizh~[2005], \S3.2 that the system can be reduced to the
known \textit{constrained least squares problem} (Lawson and Hanson~[1974],
Golub and Van Loan~[1989], Kahaner et al.~[1989]), which efficient solution is
embedded to the powerful computing environment MatLab and some other modern
systems.

\end{document}